# Modelling Pulsating Stars

Philip Masding (philmasding2024@gmail.com) Robin Leadbeater

British Astronomical Association

## Subject Keywords

Hydrodynamic codes; radial velocity measurement; spectroscopy; pulsating stars; SZ Lyn

## Abstract

Pulsating stars have been studied using non-linear hydrodynamic codes since the pioneering work of Robert Christy in the 1960's. Modern codes include improvements such as allowing for convection but there is a penalty in terms of computation speed and for some stars convection is not significant. In this work a new version of the Christy program has been developed which can run hundreds of star models to convergence in a day or two of computer time. This allows overall patterns of behaviour to be studied and suitable models for individual case stars to be identified.

The star SZ Lyn was chosen as a test case for the model. Light curve and radial velocity data were obtained for this star using amateur equipment.

A run of 625 parameter sets (mass, luminosity, effective temperature and hydrogen fraction) identified the best fit parameters for SZ Lyn. Model results show a good fit to the observed data in terms of amplitude, period and shape of the light and velocity curves.

In this paper we report on the model developed by PM and radial velocity observations by RL.

## Introduction

Pulsating variable stars are very important in astronomy particularly because the period luminosity relationship is vital for determining distances. The first Cepheid was discovered by John Goodricke in 1784 but little was known about the fundamental mechanism of variation until the early 20$^{th}$ century. Eclipses by binary companion stars were suggested but did not explain the shape of the light curves. This led to the proposal by Shapley that the variability was due to intrinsic pulsation of the star (Shapley, 1914). Beginning in 1918 the scientist and mathematician Arthur Eddington developed the heat engine theory to explain the pulsation (Eddington, 1918).

With their importance in distance determination, as exemplified by Hubble using Cepheids to establish that the Andromeda galaxy was a separate star system to the Milky Way, there was an obvious incentive to reproduce the behaviour of real stars using Eddington's theory. Unfortunately, the equations cannot be solved analytically and linear solutions can only explain the amplitude and period of the brightness curves rather than their shape. However, with the advent of powerful computers in the 1960's and the numerical mathematical techniques developed during the Manhattan project in the second world war it became possible to calculate a solution based on Eddington's theory. The pioneer in this area was Robert Christy. His work showed that it was possible to predict many of the complex features seen in variable star brightness and radial velocity curves (Christy, 1964).

The motivation of this work is to develop a modern version of Christy's code and exploit the speed of modern computers to run hundreds of trial models to match observations. Amateur astronomers have a long history of observing variable star light curves but in recent years it has become possible to observe the radial velocity as well using a spectrograph on a small telescope. The combination of a model which can run quickly on a personal computer and observational data to go with it means that amateurs can contribute to the understanding of pulsating stars.

## The mechanism of variability

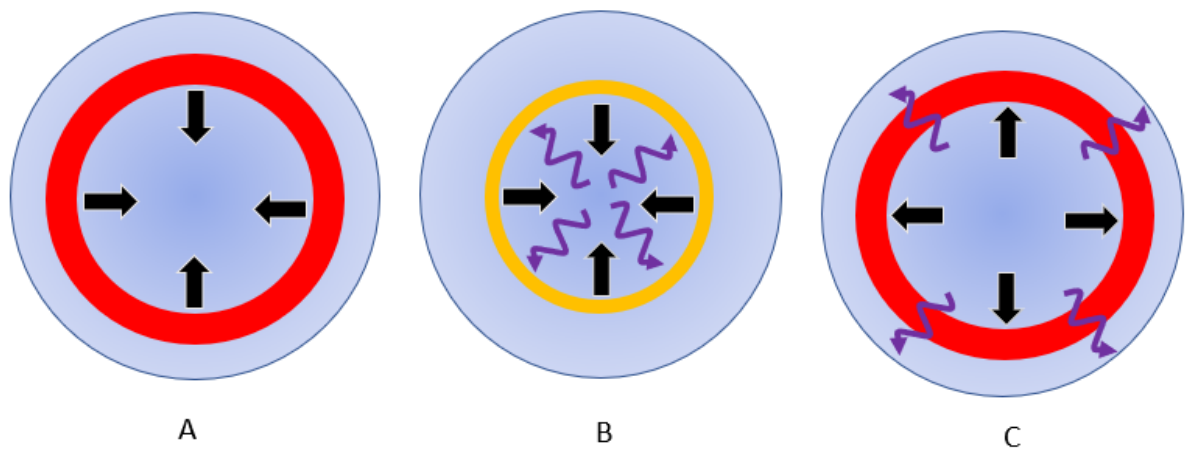

*Figure 1 Illustration of Eddington's heat engine theory*

A simple illustration of Eddington's theory is shown in Figure 1. In 1A a zone of a star is not in hydrostatic equilibrium such that the gravity force exceeds the supporting pressure so the zone falls. As it falls it is compressed becomes denser and more opaque trapping radiation in the zone below. Eventually at 1C the trapped radiation increases the pressure below the zone so much that it is forced upwards where upon it becomes relatively transparent allowing the radiation to escape and the cycle will repeat.

A problem with the initial version of the theory is that as the zone falls and is compressed it becomes denser and hotter. Although increasing density is associated with higher opacity, increasing temperature causes lower opacity and the temperature effect will predominate. An explanation was provided by Zhevakin (Zhevakin, 1953) who showed that in some stars temperatures at critical depths are just right to cause ionisation of hydrogen and helium. This ionisation absorbs energy without a temperature rise leaving the increasing density to raise opacity. This is the so-called kappa mechanism which explains why only some stars pulsate.

## The computer model

By the 1960's computers had become powerful enough to calculate a numerical solution to the pulsating star theory. Also, new methods from the atom bomb project could now be used to solve the equations for gases under extreme conditions. One of the people who had worked on the bomb project, Robert Christy, had by this time moved to Caltech and he developed a model using the IBM 7090 computer (Figure 2) at the university.

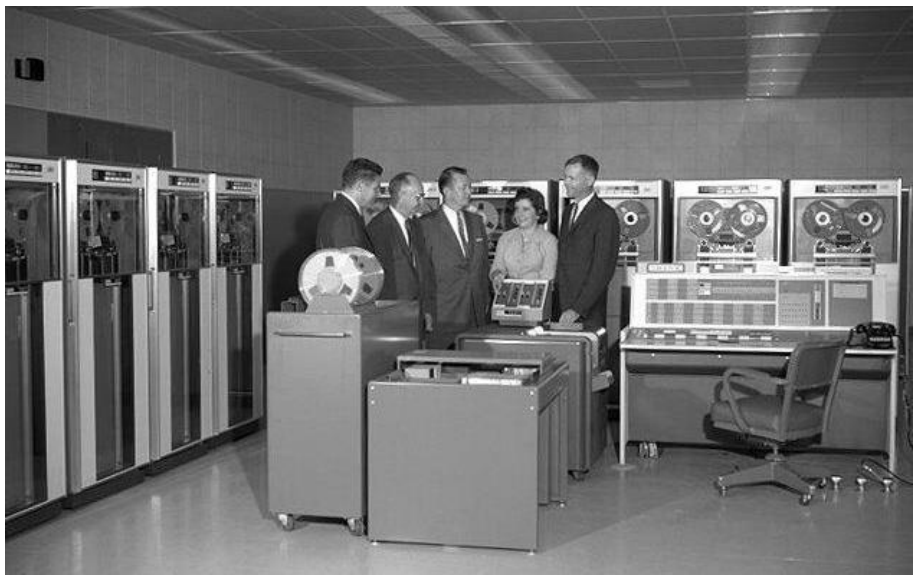

*Figure 2 An IBM 7090 computer in the 1960's (credit NASA)*

In this project a new version of Christy's code was developed in C++ to take advantage of the speed of a modern PC which is typically 10,000 times faster than the IBM7090.

This type of model is 1 dimensional which means that the star can pulsate in the radial direction only. In order to allow this the star is divided up into a number of mass zones or shells from the surface down to a depth where any pulsation is negligible which is typically 15% of the total radius. The inner boundary condition is then zero velocity and a constant radiation flux from the nuclear reactions in the core. A simplified schematic of the new code is shown in Figure 3.

There are 3 main parts to the solution beginning with the static solver. Solving the static model requires finding a set of temperatures and zone radiuses which achieve hydrostatic equilibrium or a balance of gravity and pressure forces. The other one-off calculation is to create a 2D look up table of properties of the gas mixture at various temperature and densities. This calculation requires a solution of the Saha equations to determine the state of ionisation of the hydrogen, helium and metal mixture.

In the dynamic stage the model steps forward through time with a variable step length dependent on the Courant condition or time for sound to cross the narrowest zone. Each cycle begins by solving the energy balance for the temperature using an iterative technique. The velocity and radius are then updated by explicit numerical integration.

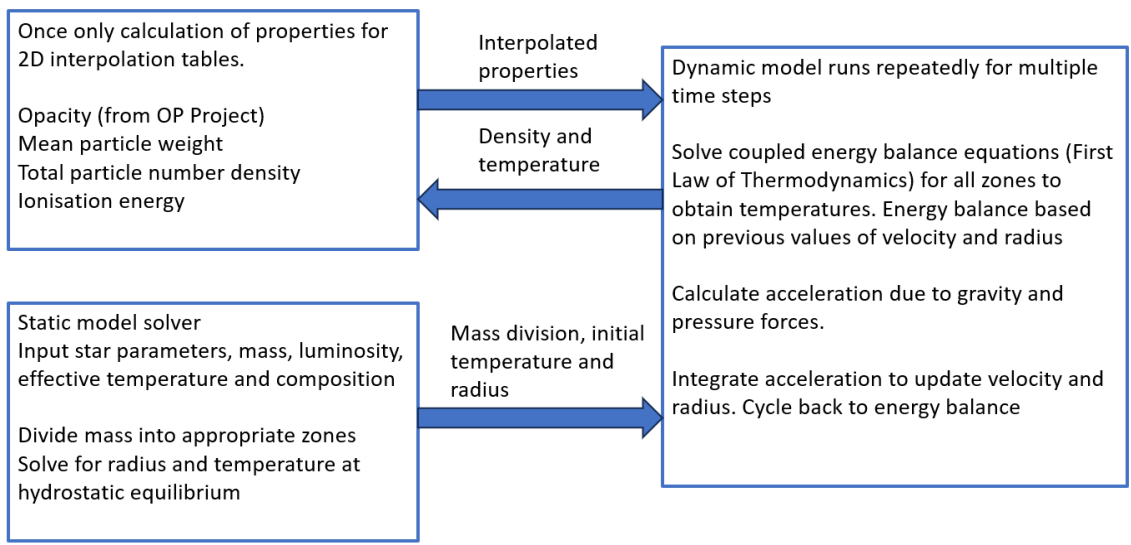

*Figure 3 The model structure*

## Observations for a test star SZ Lyn

In order to test the model, the authors obtained light curve and radial velocity data for the high amplitude $\delta$ Scuti star SZ Lyn. This star is conveniently placed for northern hemisphere observers in the spring and has a short 2.892 hour period which means all observations can be obtained in one night. At magnitude 9.08 to 9.72 it also bright enough to allow the radial velocity to be measured.

Data for the light curve was obtained unfiltered using an 80mm refractor and Atik 313+ camera. These observations were reduced using standard techniques. The radial velocity measurements are more involved and are the subject of the next section.

## Measurement of radial velocity

The star-centric radial pulsation velocity data for SZ Lyn used in this paper were derived from radial velocity measurements on spectra recorded using a LHIRES III spectrograph (resolving power R =7000) mounted on a 0.28m aperture telescope. The wavelength covered was 443-458 nm and the exposure time for each measurement was 600s. To minimise the effects of any drift in the spectrograph, the wavelength calibration was based on the average of calibration lamp spectra taken before and after each star spectrum.

For SZ Lyn, 34 spectra were recorded during the night of 7-8[th] March 2023 covering 2 complete pulsation cycles. The signal/noise ratio in each spectrum was typically 35 per resolution interval. The spectra were individually reduced using ISIS software [isis-software (astrosurf.com)](). Typical spectra recorded around maximum and minimum radial velocity are shown in Figure 4. All spectra are available to view and download from the British Astronomical Association Spectroscopy Database.

[https://britastro.org/specdb/](https://britastro.org/specdb/)

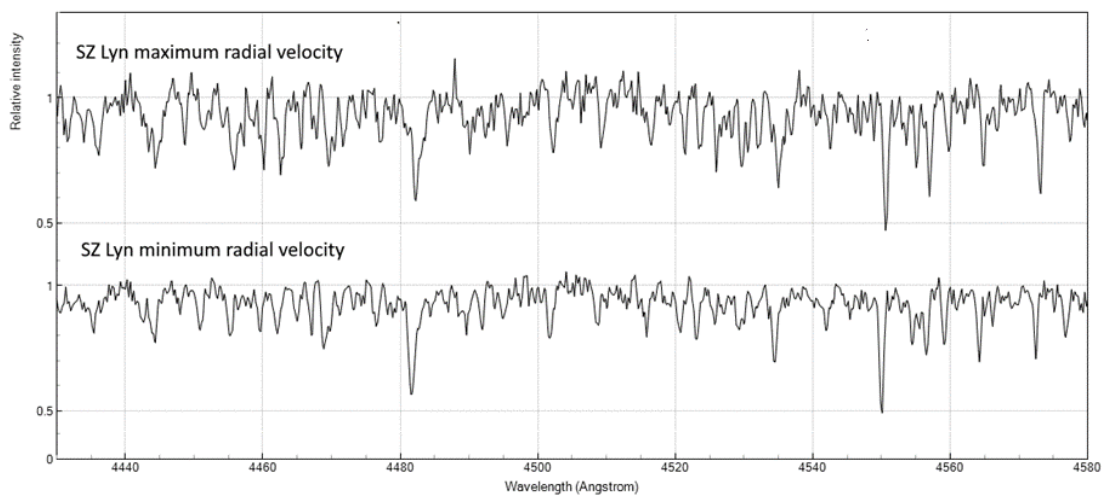

Figure 4 Spectra of SZ Lyn

The relative radial velocities were measured by the cross-correlation method, using the tool in ISIS. The first spectrum in the series was used as the reference template. The velocities and observation times were then heliocentric corrected.

The data were folded into a single cycle using a period 0.12053 day for SZ Lyn as published in the International Variable Star Index. [The International Variable Star Index (VSX) (aavso.org)]() and the phase referenced relative to the time of maximum brightness. The velocity offset (due to the motion of the star and the arbitrary choice of reference spectrum) was removed. The resulting measured radial velocities are plotted in Figure 5

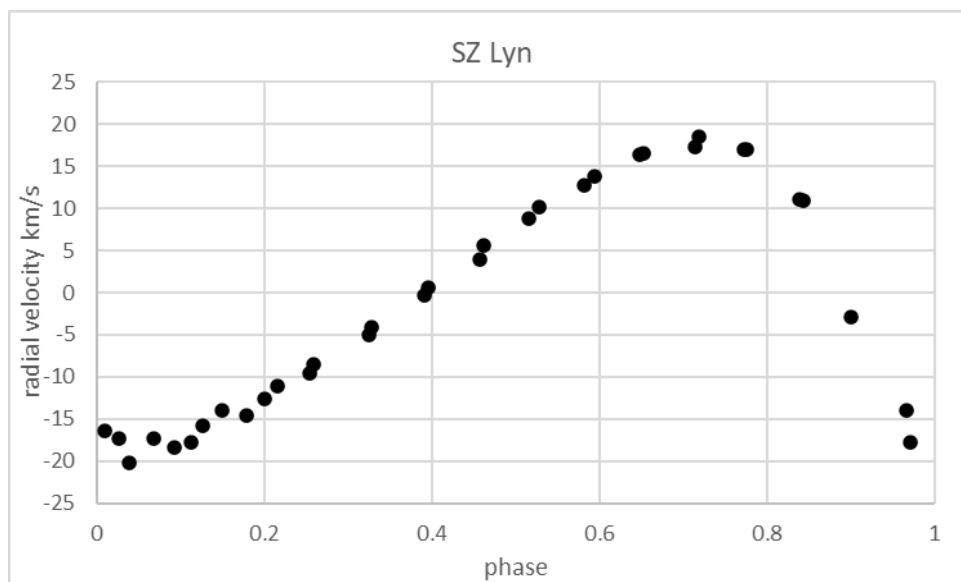

*Figure 5 Radial velocity for SZ Lyn*

A projection factor was applied to account for the fact that the measured velocity is an average over the stellar disc of the component of the star-centric radial pulsation velocity in our direction. (This component reduces towards the limb and hence the average measured velocity is smaller than the pulsation velocity.) This factor also includes the effect of limb darkening among other factors and its precise value remains of some debate and may to some degree be star specific (Borgniet, 2019). Here we have adopted a value of 1.35 with an uncertainty of ~5%.

Finally, the sign of the velocity was reversed to change the origin of the reference coordinates from heliocentric to star centric.

The stochastic uncertainty of the individual measured pulsation velocities is 2 km/s for SZ Lyn, estimated from the residuals to a quadratic fit over the falling part of the pulsation velocity curve. There is an additional systematic uncertainty arising from the uncertainty in the value of the projection factor.

Note:

SZ Lyn is a binary system with a period of 3.1 years and so the orbital motion will have an effect on the observation times and measured radial velocities and any light from the secondary would be included in the brightness measurements. (Moffett, et al., 1988). The timescale covering these observations was sufficiently short compared with the orbital period such that effect on the timings and radial velocity are negligible and were not corrected for. There is no evidence of the secondary in the visible spectrum which implies the secondary is at least 2 magnitudes fainter. (Bardin & Imbert, 1984). A 2-magnitude fainter secondary would reduce the observed magnitude range by 0.09. Here we have assumed no significant contribution from the secondary in the brightness measurements.

## Model results

The model requires a number of key parameters to define a star. For SZ Lyn suitable literature values are $7200K < T_e < 7800K$, $0.91M_\odot < M < 1.74M_\odot$ and , $2.3R_\odot < R < 2.87R_\odot$ (Adadduriya, et al., 2020 ). In order to find the best fit to the observational data the model was run with 5 values of each parameter spanning the quoted range of uncertainty. In addition, the composition was varied with 5 values of hydrogen fraction in the range $0.6 < X < 0.73$ at fixed Z=0.015. As a result, the total number of cases computed was 625. The model used a 74-zone envelope with 20km/s initial velocity imposed at the surface.

Best fit physical parameters from all these cases are $T_e = 7500K$, $R = 2.81R_\odot$, M=1.57$M_\odot$ and $X = 0.61$. With these values the model result show excellent agreement with observations made by the authors in March 2023.

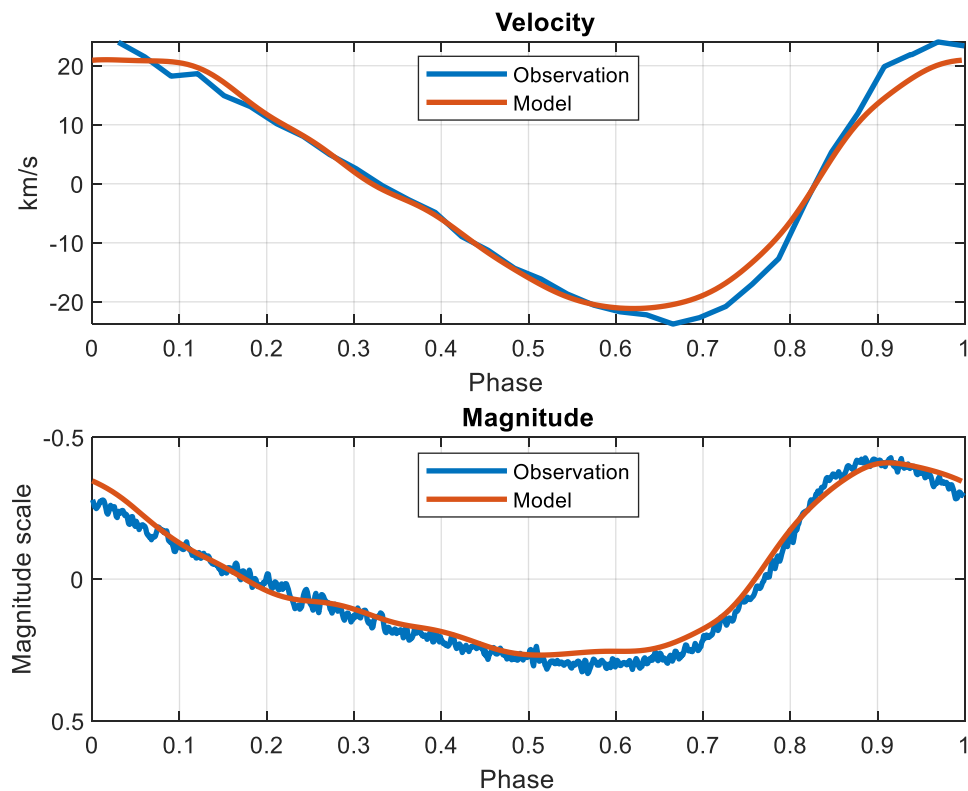

*Figure 6 Model and Observation for SZ Lyn*

## Inside the star

Observations are restricted to what happens at the surface of the star, but with the model we can predict what is happening throughout the envelope. In Figure 7 we see the velocity of the zones, with yellow colours indicating positive velocity as the star expands and blue indicating negative velocity as it contracts. Notice that significant velocity only extends down to about 75% of the total radius.

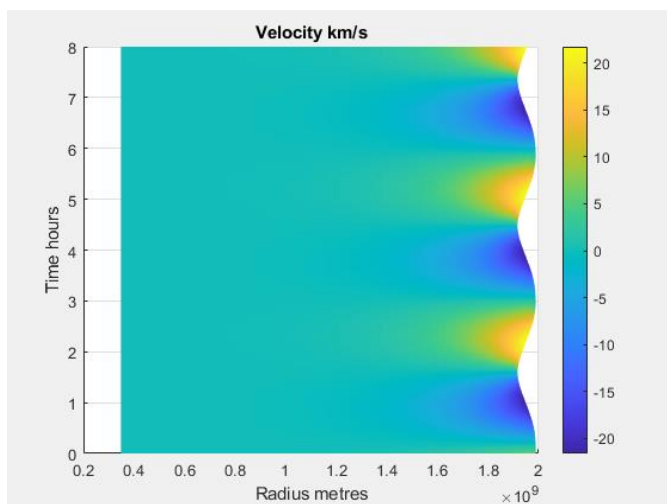

*Figure 7  Radial velocity inside the star*

Figure 8 shows the variation of HeII as a fraction of total helium. As such the bright yellow band represents 100% HeII and this band is broadest at maximum radius. The transition from HeI to HeII always takes place over a short distance near the surface whereas the transition from HeII to HeII is at a deeper level and takes place over a larger distance.

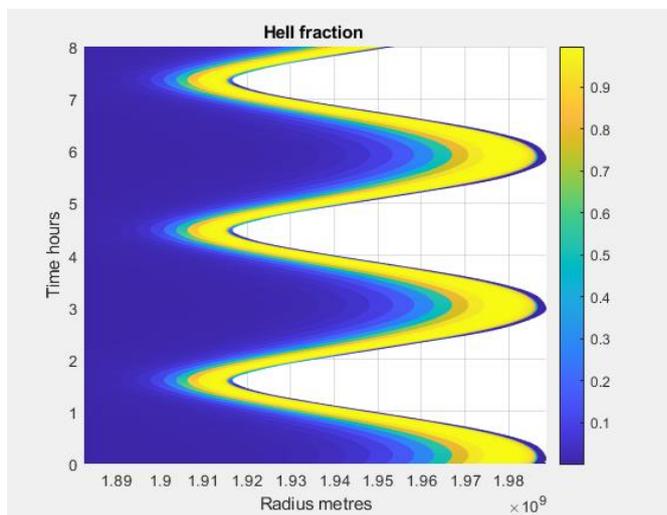

*Figure 8 Variation of HeII fraction inside the star*

The variation of opacity within the star has interesting patterns as shown by Figure 9. On this scale the cycling of hydrogen ionisation is poorly resolved since it takes place so close to the surface. In deeper zones the variation in opacity due to helium and the z-bump leads to complex cycling.

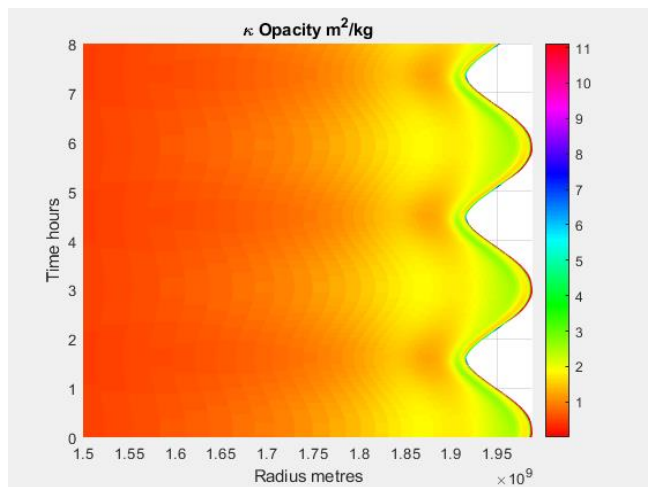

*Figure 9  Opacity*

## Conclusions

This work has demonstrated that a new implementation of the Christy hydrodynamic code can run extremely fast on a modern computer. This speed opens up the possibility to test hundreds of sets of the basic input parameters of mass, luminosity, effective temperature and composition and obtain a best fit model for observed light and radial velocity curves for a given star.

Amateur astronomers have a long history of observing light curves but in recent years spectrographs such as the LHIRES III have become available which are capable of measuring the radial velocity of stars down to at least magnitude 10 using telescopes of around 0.3m aperture. This means that amateur data can be used to fully test the model.

We obtained observational results for SZ Lyn as a test case for the model. This star is relatively bright and has a short period making it ideal to observe. In this case we screened several hundred parameter sets and found a set which very accurately fits the observed data and is within error bounds for each parameter as quoted in the literature.

In future work it is hoped to observe and fit models to more stars. The original Christy model does not include convection which does limit its applicability so a future plan is to include a simple representation of convection in the new code.

## Acknowledgements

PM would like to thank Joyce Guzik for her encouragement in submitting this work to the AAVSO conference and also thank Edward Masding for his help with C++ coding and building a fast computer.

This research has made use of the International Variable Star Index (VSX) database, operated at AAVSO, Cambridge, Massachusetts, USA.

# References

Adadduriya, J. et al., 2020 . Asteroseismology of SZ Lyn using multiband high time resolution photometry from ground and space. *Mon Not R Astron So,* 502(1).

Borgniet, A., 2019. Consistent radial velocities of classical Cepheids from the cross-correlation technique. *Astronomy & Astrophysics,* p. 631.

Christy, R. F., 1964. The Calculation of Stellar Pulsation. *Rev Mod Phys 36,* p. 555.

Eddington, A. S., 1918. On the pulsation of gaseous stars and the problem of the Cepheid variables. *Monthly notes of the Royal Astronomical Society ,* Volume 79, pp. 2 - 22.

Moffett, T. J. et al., 1988. Orbital and photometric properties of SZ Lyncis. *The Astronomical Journal,* Volume 95.

Shapley, H., 1914. On the Nature and Cause of Cepheid Variation. *Astrophysical Journal ,* Volume 40, pp. 448-465.

Zhevakin, S. A., 1953. Theory of Cepheids. *Astron J Sov. Union.*